\documentclass[12pt]{article}
\usepackage{graphicx}
\usepackage{mathptmx,mathrsfs}
\usepackage{amsfonts,amsmath,amssymb,amsthm}
\usepackage{indentfirst}
\usepackage{cite}
\usepackage{hyperref}

\numberwithin{equation}{section}
\begin{document}

\title{Nonperturbative Aspects of the two-dimensional Massive Gauged Thirring Model}
\author{R. Bufalo$^{1}$\thanks{%
rbufalo@ift.unesp.br}~ and B.M. Pimentel$^{1}$\thanks{%
pimentel@ift.unesp.br}~ \\
%EndAName
\textit{{$^{1}${\small Instituto de F\'{i}sica Te\'orica (IFT) - Universidade Estadual Paulista (UNESP)}}} \\
\textit{\small Rua Dr. Bento Teobaldo Ferraz 271, Bloco II, 01140-070, S\~ao Paulo - SP - Brazil}\\
}
\maketitle
\date{}

\begin{abstract}
In this paper we present a study based on the use of functional techniques on the issue of
insertions of massive fermionic fields in the two-dimensional massless Gauged Thirring Model.
As it will be shown, the fermionic mass contributes to the Green's functions in a surprisingly
simple way, leaving therefore the original nonperturbative nature of the massless results still
intact in the massive theory. Also, by means of complementarity, we present a second discussion
of the massive model, now at its bosonic representation.
\end{abstract}

\newpage

\section{Introduction}

Exactly solvable two-dimensional field theories \cite{5} have a non-trivial and highly interesting
structure, and by possessing a surprising richness of nonperturbative phenomena were recognized as
serving as a laboratory for studying properties of realistic theories, since they share properties
such as confinement, screening of charge, chiral symmetries, etc, with gauge theories. For instance,
the simplest two-dimensional field theory, the well-known Schwinger model, massless quantum
electrodynamics, has been explored exhaustively along the years due its resemblance with $QCD_{4}$.
Many studies were performed also in the massive case due to its attractive properties such as the
formation of bound states, chiral condensate, nontrivial vacuum structure, etc \cite{22}, and also
in a noncommutative space-time \cite{27}. Although there is an intensive investigation in two-dimensional
theories with massive fermions, none of them are concerned in analyzing an important property, the
change in the behavior of the nonperturbative character of its massless solution. Therefore, we have
as the main aim of this paper as to look carefully to this issue \cite{24} at the general Gauged
Thirring Model scenario. We wish to attain with this study be able to answer (at least, looking for
additional insight in the contrasting behavior between the massless and massive solution) the following
question: whether, or under which constraints, is it possible to preserve the original nonperturbative
characteristic of a massless fermionic model's solution when a mass term is introduced? And for our
surprise, at least in the path-integral approach, and for the fundamental Green's functions of this
particular model as well, the present study and its outcome reveal that the effects of introducing
a fermionic mass term are traced out directly in the fermionic Green's functions. Moreover, the
massive contribution is remarkably resumed in a simply way, leaving intact the nonperturbative
character of the solution of the massless model.

As it has been initially proposed, the Gauged Thirring model (GTM) is a generalization of the usual
Thirring model; where the local gauge symmetry was implemented by using the Hidden Local Symmetry
technique \cite{11,15}. However, one of the most interesting features of the GTM was pointed out by
Kondo in his proposal \cite{12}; where it is introduced not only one auxiliary scalar field but also
a kinetic term for the gauge field in the Thirring model. Within this proposal, the GTM, at classical
level and strong coupling regime, behaves as the electrodynamics with fermions (for $%
g\rightarrow\infty$) or, as a fermionic current-current self-interaction field theory
(for $e^{2}\rightarrow\infty$). In the latter case, at $(1+1)$-dimensional spacetime, the first
model is known as the Schwinger model (SM), and the second one as the Thirring model (TM).
Furthermore, it was recently proved that the equivalence between the Gauged Thirring model
and the Schwinger and Thirring models at $(1+1)$-dimensions is also present at quantum
level \cite{14}. Such equivalence was proved through a study of a nonpertubative quantization
and by further analysis of respective Green's functions and Ward-Fradkin-Takahashi
identities as well.

An interesting framework to deal with massive fermionic theories is to look its bosonic
formulation. The bosonization of the $(1+1)$-dimensional field models has been widely
investigated along the years, and has its historical roots in Klaiber's work on Thirring
model \cite{1}, and at Lowestein and Swieca's work on Schwinger model \cite{2}. Since
Coleman's proof, regarding the equivalence between the massive Thirring and sine-Gordon
models \cite{6}, the concept of bosonization has been claimed as being a remarkable tool
to obtain nonperturbative information of $(1+1)$-dimensional theories. The bosonization
prescription is now well understood and rigorously established at both operator and
path-integral frameworks \cite{8,20,26}, and also in different spacetime dimensionality
\cite{7}. Nowadays the bosonization of two-dimensional models at finite temperature also
have been received attention \cite{16}. These features will be explored here to perform
the bosonization prescription for the massive Gauged Thirring model (MGTM). And, once
the bosonic prescription is established to the massive theory, there are several and
relevant phenomena to be exploited; e.g., we have interest in analyzing the solutions
and thermodynamics of the present model at curved space-time \cite{32}.

Therefore, in this paper, we shall present a detailed study of the two-dimensional massive
Gauged Thirring model from the point of view of the path-integral framework. The main focus
of the present discussion will consist in analyzing the behavior of the massless GTM
in face of a perturbative expansion of the mass parameter. To this goal, we will make use
of the mass insertions \cite{5,24} on the massless model's generating functional. Also, as
it was mentioned earlier, we will show that the effects of the fermionic mass will appear
in such a way that the nonperturbative nature of the massless results \cite{14} will be
preserved in the presence of the massive fermionic fields.

The paper is organized as it follows: In the Sect.\ref{sec:1.1}, our main task will be to
deal with the fermionic mass insertion into the generating functional of the $(1+1)$-dim.
massless Gauged Thirring model, by making use of the functional derivative techniques and
treating the mass parameter perturbatively. As it will be shown, the fermionic mass
contributes only to the fundamental Green's functions in a surprisingly simple way, leaving
the nonperturbative phenomena of the massless results still intact. In the Sect.\ref{sec:1},
we approach the massive theory by a different perspective, now finding the bosonic version
of the massive GTM. In deriving this bosonic representation, we will make use of the
path-integral bosonization prescription. At last, we present our final remarks and prospects in
the Sect.\ref{sec:3}.

\section{Massive Gauged Thirring model}

\label{sec:1.1}

Due to the non-trivial and rich structure, and exact nonperturbative outcome of the massless
theory, we are strongly motivated to investigate whether these phenomena (also other ones) are
preserved when massive fermionic fields are considered. Although both subjects that we wish to
discuss here are unrelated, we want to provide a first study and interpretation on the massive
features in different frameworks to thus be able to detect the deviations attached to the mass
effects, and also to discuss how plausible this analysis is whether we are interested in
preserving the nonperturbative aspects of the solution of a given two-dimensional field theory.

Nevertheless, we start by reviewing the path-integral approach applied to the MGTM, then to
discuss next explicitly the details of our analysis. The dynamics of MGTM is described by the
following Lagrangian density \cite{12,14}:
\begin{equation}
\mathcal{L}=\bar{\psi}\left( i\hat{\partial}+\hat{A}\right) \psi -m\bar{%
\psi}\psi +\frac{1}{2g}\left( A_{\mu }-\partial _{\mu }\theta \right) ^{2}-%
\frac{1}{4e^{2}}F_{\mu \nu }F^{\mu \nu },  \label{eq a.1}
\end{equation}%
which is invariant under the local gauge transformations:%
\begin{equation}
\psi ^{\prime } =e^{i\sigma \left(x\right) }\psi  , \quad A_{\mu }^{\prime }=A_{\mu }
+\partial _{\mu }\sigma \left(x\right)  ,\quad \theta ^{\prime } =\theta  +\sigma \left(x\right) .  \label{eq a.2}
\end{equation}%
The $\theta $-field was introduced into the Lagrangian \eqref{eq a.1} following the
St\"{u}ckelberg procedure in order to incorporate the local gauge symmetry.
Furthermore, the system is defined into a Minkowskian $(1+1)$-dimensional spacetime
with metric: $\eta _{\mu \nu }=diag\left( 1,-1\right)$, and with the following
representation for the Dirac $\gamma $-matrices: $ \gamma _{0}=\sigma _{1},~
\gamma _{1}=-i\sigma _{2},~\gamma _{5}=\gamma _{0}\gamma _{1}=\sigma _{3}$, and
with the $\gamma $-matrices satisfying the algebra: $ \{\gamma _{\mu },\gamma _
{\nu }\}=2\eta _{\mu \nu }$, and the identity: $\gamma _{\mu }\gamma_{5}=\epsilon
 _{\mu \nu }\gamma ^{\nu }$, where $\epsilon^{\mu \nu }=\epsilon _{\mu \nu }$ and
$\epsilon _{01}=-\epsilon _{10}=1$. Now, one may obtain the generating functional
for the theory from the Lagrangian density \eqref{eq a.1} at $R_{\xi }-$gauge:
\begin{equation}
R_{\xi }=\partial _{\mu }A^{\mu }\left( x\right) +\frac{\xi }{g}\theta
\left( x\right) .  \label{eq a.3.1}
\end{equation}%
such choice is known to decouples the $\theta $-field from the other fields, then the
generating functional is written as \cite{14}
\begin{align}
\mathscr{Z}\left( \eta ,\bar{\eta},J^{\mu },C^{\mu },K\right) =\int D\theta
DA_{\mu }D\bar{\psi}D\psi \det \left\vert \square +\frac{\xi }{g}\right\vert \exp \left( i\int d^{2}x\left[ \mathcal{L_{\psi ,A}}+\mathcal{L_{\theta }}+%
J.\Phi+K\theta\right] \right) , \label{eq a.3}
\end{align}%
with the quantities defined as the following:
\begin{equation}
\mathcal{L_{\psi ,A}} =\bar{\psi}\left( i\hat{\partial}+\hat{A}
-m\right) \psi -\frac{1}{4e^{2}}F_{\mu \nu }F^{\mu \nu }+\frac{1}{2g}A^{\mu }A_{\mu }
-\frac{1}{2\xi }\left( \partial _{\mu }A^{\mu }\right) ^{2},
\end{equation}
and
\begin{equation}
\mathcal{L_{\theta }} =\frac{1}{2g}\left( \partial _{\mu }\theta \right)
\left( \partial ^{\mu }\theta \right) -\frac{\xi }{2g^{2}}\theta ^{2},
\end{equation}
and the short notation for the fermionic, electromagnetic and fermionic current \footnote{Here
we define the fermionic current as $j_\mu = \bar{\psi}\gamma_\mu \psi$.} source terms
\begin{equation}
J.\Phi=\bar{\eta}\psi +\bar{\psi}\eta +J_{\mu }A^{\mu } +C_{\mu }j^{\mu }.
\end{equation}
The majority of the interest in two-dimensional field theories relies mainly on the
nonperturbative character of its solutions. Unfortunately, such analyses are always limited
to the case of massless fermionic fields and few papers only are devoted in dealing carefully
with the case of massive fields, and most of them are only concerned to the first leading term and its
possible deviation from massless results \cite{30}. But, as matter of illustration, considering
for instance, Eq.\eqref{eq a.3} in an $\omega$-dimensional spacetime and the fermionic fields $\psi,~\bar{\psi}$
interacting with vector fields $B^{a}_{\mu}$ (here we have included all the non-fermionic fields), we have
after integrating over these fermionic fields
\begin{align}
\mathscr{Z}\left( J\right) =&\int DB^{a}_{\mu } \det \left( i\hat{\partial}-m+\hat{B}^{a}\right) \exp
\left( i\int d^{\omega}x\left[ \mathcal{L}_{B^{a}}+J.B \right] \right) \notag \\ 
&\times \exp \left( -i\int d^{\omega}xd^{\omega}y\bar{\eta}\left( x\right) \mathfrak{S}
\left( x,y;m;B\right)\eta\left( y\right) \right) \label{eq 3.2a},
\end{align}
where the massive external field fermionic propagator satisfies: {\small $\left( i\hat{\partial}
-m+\hat{B}\right) \mathfrak{S}\left( x;m;B\right) =\delta \left( x\right) $}. There are two
equivalent ways to solve the above functional determinant: either we may consider a perturbative
expansion in the vector field $B^{a}_{\mu}$ as
\begin{align}
\det \left( i\hat{\partial}-m+\hat{B}^{a}\right)&= \det \left( i\hat{\partial}-m\right)\det\left( 1+
\hat{B}^{a}S\left( x,y;m\right)\right), \notag\\
&= \det \left( i\hat{\partial}-m\right)\exp \left[tr \ln\left( 1+\hat{B}^{a}S\left( x,y;m\right)\right) \right] \label{eq 3.1},
\end{align}
where $S\left( x;m\right)$ is the massive free fermionic propagator; moreover, we can also write an
equivalent expansion in $m$ as
\begin{align}
\det \left( i\hat{\partial}-m+\hat{B}^{a}\right)= \det \left( i\hat{\partial}+B^{a}\right)
\exp \left[tr \ln\left( 1-m \mathfrak{S}\left( x,y;B\right)\right) \right] \label{eq 3.2},
\end{align}
where $\mathfrak{S}\left( x;B\right)$ is the massless external field fermionic propagator. The
form in Eq.\eqref{eq 3.1} is usually suitable for perturbative calculation in the gauge
interaction in any spacetime dimensionality. Nevertheless, the expression \eqref{eq 3.2} is
mostly suitable in the case of the two-dimensional field theory, because in this framework we
can evaluate exactly the determinant $\det \left( i\hat{\partial}+B^{a}\right)$, obtaining a rich
and interesting nonperturbative outcome in the first determinant but treating perturbatively the
second massive functional. \footnote{For that matter we refer to the Ref.~\cite{31} and Section 4.9
of the Ref.~\cite{5} for a more detailed discussion.} Besides, as it is well-known the full
nonperturbative character of $\left( 1+1\right) $-dimensional models is encoded into the
exactly computation of the massless fermionic determinant.

However, we will follow a third way to solve the functional fermionic integration which resembles
somehow the second way, but it avoids the need in evaluating the (second) massive determinant by making
use of functional techniques before performing the fermionic integration, instead we shall need
to evaluate a perturbative series of functional derivatives only. Therefore, in order to
implement this idea, we will proceed as it is usually done in perturbative calculation in the path-integral
approach (when one substitutes the interacting term of Lagrangian by its Fourier transform in terms of the functional
derivatives of sources $\mathcal{L}_{int}\left[\varphi(x)\right]\rightarrow \mathcal{L}_{int}
\left[\frac{\delta}{i\delta J(x)}\right]$), we will substitute the mass term as in terms of functional derivatives
\begin{align}
-m \int d^{2}z \bar{\psi}\left( z\right)\psi \left( z\right) \rightarrow m\int d^{2}z
\frac{\delta ^{2}}{i\delta \eta \left(z\right) i\delta \bar{\eta}\left( z\right) }.
\end{align}
within this approach we will deal with the fermionic mass parameter perturbatively on the generating
functional \eqref{eq a.3} by means of functional techniques. With this analysis of the massive theory
we hope to provide a wealth discussion on the subject as well as to demystify the fact that the
nonperturbative character of such massless fermionic theories can also be present at a massive theory.
The next step consists in computing the (simplest) fundamental Green's functions and analyzing how is
their dependence in powers of the mass parameter and, moreover, its behavior deviation in relation
from the massless ones, in the sense of which properties are preserved (broken) in this generalization
\cite{5}. As it was already discussed in the Ref.\cite{14}, the $\theta $-field is decoupled from
the other fields, therefore we can work only with the following generating functional:
\begin{equation}
\mathscr{Z}\left( J\right) =\int DA_{\mu }D%
\bar{\psi}D\psi \exp \left( i\int d^{2}x\left[ \mathcal{L_{\psi ,A}}+J.\Phi\right] \right) .\label{eq a.4}
\end{equation}
Nevertheless, proceeding as explained above, the resulting expression after performing the fermionic integration
may be written in terms of the gauge field and the fermionic sources and its derivatives (mass contribution):
\begin{align}
\mathscr{Z}\left( J\right)&=\int DA_{\mu
}\det \left( i\hat{\partial}+\hat{A}+\hat{C}\right) \exp \left[ im\int d^{2}z\frac{\delta ^{2}}{i\delta \eta \left(
z\right) i\delta \bar{\eta}\left( z\right) }\right] \notag \\
&\times \exp \Big( i\int d^{2}x \Big[ \frac{1}{2g}A^{\mu }A_{\mu}-\frac{1}{4e^{2}}F_{\mu \nu }F^{\mu \nu }
-\frac{1}{2\xi }\left( \partial _{\mu }A^{\mu }\right) ^{2}+J_{\mu }A^{\mu }\Big] \Big) \notag \\
& \times \exp \left( -i\int d^{2}xd^{2}y\bar{\eta}_{B}\left( x\right) \mathfrak{S}_{BD}\left( x,y;A+C\right) \eta
_{D}\left( y\right) \right) .  \label{eq a.5c}
\end{align}%
It is important to emphasize that the difference between the expressions \eqref{eq 3.2a}, \eqref{eq 3.2}  and \eqref{eq a.5c}
is mainly where the mass effects are present: in the usual approach, Eqs.\eqref{eq 3.2a} and \eqref{eq 3.2}, the mass appears
in both functional determinant and at the complete massive Dirac propagator $\mathfrak{S}\left( x;m;B\right)$, while in our proposal,
Eq.\eqref{eq a.5c}, it appears as depending only of the derivatives of fermionic sources. Moreover, throughout the calculation
is being used the following definition for grassmannian derivative:%
\begin{equation}
\frac{\delta }{\delta \zeta _{A}}\left( \zeta _{A_{1}}\zeta _{A_{2}}\right)
=-\delta _{AA_{1}}\zeta _{A_{2}}+\zeta _{A_{1}}\delta _{AA_{2}};  \label{eq a.8.1}
\end{equation}%
the capital letters stand for spinorial indices, and we also have that the massless external
field fermionic propagator $\mathfrak{S}\left( x,y;A\right)$ has the following well-known solution
in two-dimensions \cite{5,14,45}:%
\begin{equation}
\mathfrak{S}\left( x,y;A\right) =\exp \left[ -i\int d^{2}wA_{\mu }\left( w\right)
s^{\mu }\left( w,x,y\right) \right] S\left( x-y\right) ,
\label{eq a.7}
\end{equation}%
with: {\small $s^{\mu }\left( w,x,y\right) =\left( \partial _{w}^{\mu }+\gamma _{5}\tilde{\partial}
_{w}^{\mu }\right) \left( D\left(w-x\right) -D\left( w-y\right) \right) $}, where we have introduced
the notation $\tilde{v}_{\alpha }=\epsilon _{\alpha\mu }v^{\mu }$; moreover, $D$ and $S$ are the
massless free Klein-Gordon and Dirac propagator, respectively. The nonperturbative functional expression
for the massless fermionic determinant in a given external field $B$ is explicitly given by \cite{5,14,45}:%
\begin{equation}
\det \left( i\hat{\partial}+\hat{B}\right) =\exp \left( \frac{i}{%
2\pi }\int d^{2}z B_{\mu }T^{\mu \nu }B_{\nu }\right) ,  \label{eq a.8}
\end{equation}%
with the set of operators:%
\begin{equation}
L^{\mu \nu }=\frac{\partial ^{\mu }\partial ^{\nu }}{\square },\qquad T^{\mu
\nu }+L^{\mu \nu }=\eta ^{\mu \nu }.
\end{equation}%
Hence, in hands of the nonperturbative expression \eqref{eq a.8}, it follows that we can cast the
generating functional \eqref{eq a.5c} into the following form:
\begin{align}
 \mathscr{Z}\left( J\right) &=\int DA_{\mu
}\exp \left[ im\int d^{2}z\frac{\delta ^{2}}{i\delta \eta \left(
z\right) i\delta \bar{\eta}\left( z\right) }\right]   \exp \left( -i\int d^{2}xd^{2}y\bar{\eta}\left( x\right) \mathfrak{S}\left( x,y;A+C\right) \eta
\left( y\right) \right) \notag \\
&\times \exp \Big( i\int d^{2}x\Big[ \frac{1}{2}A_{\mu }W_{\xi }^{\mu \nu }A_{\nu
}+\frac{1}{2\pi }C_{\mu }T^{\mu \nu }C_{\nu } +\frac{1}{\pi }A_{\mu }T^{\mu \nu }C_{\nu } +J_{\mu }A^{\mu }\Big] \Big), \label{eq a.9} 
\end{align}%
where we had defined the differential operator $W_{\xi }$ by:%
\begin{equation}
W_{\xi }^{\mu \nu }=\frac{1}{\pi }\left( b+c\square \right) T^{\mu \nu }+%
\frac{1}{\pi }\left( b-1+\frac{\pi }{\xi }\square \right) L^{\mu \nu },
\end{equation}%
with the parameters: $b=1+\frac{\pi }{g}$ and $c=\frac{\pi }{e^{2}}$. Now, in possess of the generating
functional \eqref{eq a.9} we can derive any quantity of interest. But, as we are interested here in showing
how the nonperturbative nature of the results of the massless fermionic theory behave when a perturbative
calculation on the mass parameter is considered, we shall compute now some of the fundamental Green's functions.

\subsection{Correlation Functions}

From a simple analysis of the generating functional Eq.\eqref{eq a.9} we can see that the $m$-dependent
terms involve powers of the derivatives of the fermionic sources $\eta $ and $\bar{\eta}$. Furthermore,
the fundamental Green's functions have zero or linear dependence on the derivatives of the fermionic
sources $\eta $ and $\bar{\eta}$. Thus, it is clear that the gauge field $\mathscr{D}_{\mu \nu }^{\xi }$
and the current-current $\mathscr{J}_{\mu \nu }^{\xi }$ propagators remain unaltered in face of the
massive contribution, once they do not involve derivatives of the fermionic sources $\eta $ and
$\bar{\eta}$, i.e., they possess the same expressions as for the massless GTM \cite{14}. Hence, we now come
to the main subject of this paper, focusing all our attention into the Green's functions which do involve
$\eta $ and $\bar{\eta}$ derivatives in their evaluation. Nevertheless, in a first analysis, we shall
pay attention in the computation of fundamental functionals such as: the $2$-points fermionic function
and the two vertex functions, defined by \cite{14}:
\begin{align}
\mathscr{S}_{IK}^{\xi }\left( x-y\right) &= -\bigg. \frac{\delta ^{2}\mathscr{Z}\left( \eta ,\bar{%
\eta}\right) }{\delta \eta _{K}\left( y\right) \delta \bar{\eta}_{I}\left(
x\right) }\bigg\vert _{s=0},  \notag \\
\Gamma _{IK}^{\mu }\left( x,y;z\right) &= i\bigg. \frac{\delta ^{3}%
\mathscr{Z}\left( \eta ,\bar{\eta},J\right) }{\delta J_{\mu }\left(
z\right) \delta \eta _{K}\left( y\right) \delta \bar{\eta}_{I}\left(
x\right) }\bigg\vert _{s=0},  \notag \\
\mathscr{H}_{IK}^{\mu }\left( x,y;z\right) &=i\bigg. \frac{\delta ^{3}%
\mathscr{Z}\left( \eta ,\bar{\eta},C\right) }{\delta C_{\mu }\left(
z\right) \delta \eta _{K}\left( y\right) \delta \bar{\eta}_{I}\left(
x\right) }\bigg\vert _{s=0}.  \label{eq a.10}
\end{align}%
It can be easily seen that all these three functionals, as defined above, have the same $\eta $ and
$\bar{\eta}$ derivative structure. Moreover, such dependence can be casted, by the manipulation of
the equations \eqref{eq a.9} and \eqref{eq a.10}, in terms of the following quantity:
\begin{align}
\Phi _{IK}\left( x,y\right) &=\bigg. \frac{\delta ^{2}}{\delta \eta
_{K}\left( y\right) \delta \bar{\eta}_{I}\left( x\right) }\exp \left[ im\int
d^{2}z\frac{\delta ^{2}}{i\delta \eta \left( z\right) i\delta \bar{\eta}%
\left( z\right) }\right]  \notag \\
& \times \exp \left( -i\int d^{2}wd^{2}v\bar{\eta}_{B}\left(
w\right) \mathfrak{S}_{BD}\left( w,v;A+C\right) \eta _{D}\left( v\right) \right)
\bigg\vert _{\eta =\bar{\eta}=0},\label{eq a.11}
\end{align}%
which is exactly the same quantity computed in the Ref.~\cite{14} to the massless case, $m=0$. Therefore,
we will pay attention only to the massive features, once the subsequent calculation and further analysis
follows exactly as the one presented previously in the massless case \cite{14}. So, our main task here relies
in the explicit computation of the expression \eqref{eq a.11}. Although we are interested, initially, in
analyzing the role played by the mass parameter only in the fundamental Green's functions, we believe that
the core of this idea can be expanded to any particular interesting feature of a given two-dimensional fermionic field theory.

\subsubsection{Mass dependence calculation}

The evaluation of the several functional derivatives presented in the quantity $\Phi _{IK}\left(x,y\right) $
is lengthy, but straightforward, and we will present the main steps and further discussion on its evaluation
right above. Anyhow, we shall pay attention first to the following quantity:
\begin{equation}
\Phi \left( \eta ,\bar{\eta};m\right) =\exp \left[ im\int d^{2}z\frac{%
\delta ^{2}}{i\delta \eta \left( z\right) i\delta \bar{\eta}\left( z\right) }%
\right] \exp \left( -i\int d^{2}wd^{2}v\bar{\eta}\left( w\right)
\mathfrak{S}\left( w,v;B\right) \eta \left( v\right) \right) \notag
\end{equation}
or, explicitly, term by term
\begin{align}
\Phi \left( \eta ,\bar{\eta};m\right) =&\bigg\{1+\left( im\right) \int d^{2}x_{1}\frac{\delta ^{2}}{i\delta \eta_{D}\left( x_{1}\right)
 i\delta \bar{\eta}_{D}\left( x_{1}\right) }\notag \\
&+\frac{\left( im\right) ^{2}}{2!}\int d^{2}x_{1}d^{2}x_{2}\frac{\delta ^{2}}{i\delta \eta _{D_{1}}
\left( x_{1}\right) i\delta \bar{\eta}_{D_{1}}\left(x_{1}\right) }\frac{\delta ^{2}}
{i\delta \eta _{D_{2}}\left( x_{2}\right)i\delta \bar{\eta}_{D_{2}}\left( x_{2}\right) } +... \notag \\
&+\frac{\left( im\right) ^{n}}{n!}\int d^{2}x_{1}...d^{2}x_{n}\frac{\delta ^{2}}{i\delta \eta _{D_{1}}
\left( x_{1}\right) i\delta \bar{\eta}_{D_{1}}\left( x_{1}\right) }...\frac{\delta^{2}}{i\delta \eta _{D_{n}}
\left( x_{n}\right) i\delta \bar{\eta}_{D_{n}}\left( x_{n}\right) }+...\bigg\}  \notag \\
&\times \exp \left( -i\int d^{2}wd^{2}v\bar{\eta}_{B}\left( w\right)
\mathfrak{S}_{BF}\left( w,v;B\right) \eta _{F}\left( v\right) \right) . \label{eq a.6.1}
\end{align}%
As a matter of illustration, we present now the explicit evaluation of the first non-trivial terms of the
above expansion. Making use of the definition \eqref{eq a.8.1}, one can easily show that \footnote{In a way
to abbreviate our notation we shall write: $\mathfrak{S}_{BF}\left(w,v;B\right)= \mathfrak{S}_{BF}\left(w,v\right)$}
\begin{align}
\Phi ^{\left( 1\right) } =&\left(im\right) \int d^{2}x_{1}\frac{\delta ^{2}}{i\delta \eta _{D_{1}}\left(
x_{1}\right) i\delta \bar{\eta}_{D_{1}}\left( x_{1}\right) }\exp \left(-i\int d^{2}wd^{2}v\bar{\eta}_{B}
\left( w\right) \mathfrak{S}_{BF}\left(w,v\right) \eta _{F}\left( v\right) \right) \notag
\end{align}
and
\begin{align}
\Phi ^{\left( 1\right) } =&\left( im\right) \int d^{2}x_{1}\bigg\{ -\int d^{2}x_{2}d^{2}x_{3}%
\bar{\eta}_{B}\left( x_{2}\right)\mathfrak{S}_{BD_{1}}\left(x_{2},x_{1}\right)  \mathfrak{S}_{D_{1}F}\left(
x_{1},x_{3}\right) \eta _{F}\left( x_{3}\right) \notag \\
& +\left( -\right) \frac{1}{i}%
\mathfrak{S}_{D_{1}D_{1}}\left( x_{1},x_{1}\right)\bigg\} \exp \left( -i\int d^{2}wd^{2}v\bar{\eta}_{B}
\left( w\right) \mathfrak{S}_{BF}\left( w,v\right) \eta _{F}\left( v\right) \right) , \label{eq a}
\end{align}%
moreover, the second term
\begin{align}
\Phi ^{\left( 2\right) } &=\frac{\left( im\right) ^{2}}{2}\int d^{2}x_{1}d^{2}x_{2}\bigg\{\left( -\right)
 Tr\left[ i\mathfrak{S}\left( x_{1},x_{2}\right) i\mathfrak{S}\left( x_{2},x_{1}\right) \right] \notag \\
&+2\int d^{2}x_{3}d^{2}x_{4}\left[ \bar{\eta}_{B}\left( x_{3}\right) \mathfrak{S}_{BD_{1}}\left( x_{3},x_{1}\right)
i\mathfrak{S}_{D_{1}D_{2}}\left(x_{1},x_{2}\right) \mathfrak{S}_{D_{2}F}\left( x_{2},x_{4}\right) \eta
_{F}\left( x_{4}\right) \right]   \nonumber \\
&-Tr\left[ i\mathfrak{S}\left( x_{2},x_{2}\right) \right] \int d^{2}x_{3}d^{2}x_{4}\left[ \bar{\eta}_{B}
\left( x_{3}\right) \mathfrak{S}_{BD_{1}}\left( x_{3},x_{1}\right) \mathfrak{S}_{D_{1}F}\left(
x_{1},x_{4}\right) \eta _{F}\left( x_{4}\right) \right]   \nonumber \\
&+\int d^{2}x_{3}d^{2}x_{4}d^{2}x_{5}d^{2}x_{6}\left[ \bar{\eta}_{B_{1}}\left( x_{3}\right) \mathfrak{S}_{B_{1}D_{1}}
\left(x_{3},x_{1}\right) \mathfrak{S}_{D_{1}F_{1}}\left( x_{1},x_{4}\right) \eta_{F_{1}}\left( x_{4}\right) \right] \notag \\
&\times \left[ \bar{\eta}_{B_{2}}\left(x_{5}\right) \mathfrak{S}_{B_{2}D_{2}}\left( x_{5},x_{2}\right)
 \mathfrak{S}_{D_{2}F_{2}}\left( x_{2},x_{6}\right) \eta _{F_{2}}\left( x_{6}\right)\right] \bigg\}  \nonumber \\
&\times \exp \left( -i\int d^{2}wd^{2}v\bar{\eta}_{B}\left( w\right)
\mathfrak{S}_{BF}\left( w,v\right) \eta _{F}\left( v\right) \right) .\label{eq b}
\end{align}
By means of the solution \eqref{eq a.7}, one can show that the second term of the Eq.\eqref{eq a} is
identically null once it is proportional to $Tr[\gamma^{\mu }]=0$, the same argument is valid for the
third term of \eqref{eq b}. Actually, we can use the general result:%
\begin{equation}
Tr\left[ \#odd~of~\gamma ^{\prime }s\right] =0,
\end{equation}%
to show that the trace of any odd number of Dirac propagators is null,
\begin{align}
 Tr[\mathfrak{S}\left( x_{1},x_{2}\right) \mathfrak{S}\left( x_{2},x_{3}\right)...\mathfrak{S}
\left( x_{2N-1},x_{1}\right) ] &\propto   Tr[S\left(x_{1},x_{2}\right) S\left( x_{2},x_{3}\right) ...S\left(
x_{2N-1},x_{1}\right) ]\notag \\
& \propto  Tr\left[ \#odd~of~\gamma ^{\prime }s\right] =0.
\end{align}%
Moreover, we also can derive another result often used in the present calculation. By taking into account
the following quantity: the integral of the trace of any even number of Dirac propagators \eqref{eq a.7}, one can show that
\begin{equation}
\int d^{2}x_{1}d^{2}x_{2}...d^{2}x_{2N} Tr\left[ \mathfrak{S}\left(
x_{1},x_{2}\right) \mathfrak{S}\left( x_{2},x_{3}\right) ...\mathfrak{S}\left(
x_{2N},x_{1}\right) \right] =0;
\end{equation}%
since it is proportional to the dimensional regularization result: $\int
d^{\omega }p\frac{1}{\left( p^{2}\right) ^{N}}=0$. Nevertheless, in resume, we have
here that the first two nonvanishing terms of the expansion \eqref{eq a.6.1} are written as
\begin{align}
\Phi ^{\left( 1\right) } =&\left(- im\right) \int d^{2}x_{1}d^{2}x_{2}d^{2}x_{3}\left[
\bar{\eta}\left( x_{2}\right) \mathfrak{S}\left( x_{2},x_{1}\right)  \mathfrak{S}
\left( x_{1},x_{3}\right) \eta\left(x_{3}\right) \right] \notag \\
&\times\exp \left( -i\int d^{2}wd^{2}v\bar{\eta}_{B}\left(
w\right) \mathfrak{S}_{BF}\left( w,v\right) \eta _{F}\left( v\right) \right) ,
\end{align}
and
\begin{align}
\Phi ^{\left( 2\right) } =&\left( -im\right) ^{2}\int d^{2}x_{1}...d^{2}x_{4}\left[ \bar{\eta}\left(
x_{3}\right) \mathfrak{S}\left( x_{3},x_{1}\right) i\mathfrak{S}\left( x_{1},x_{2}\right) \mathfrak{S}\left(
x_{2},x_{4}\right) \eta\left( x_{4}\right) \right]   \nonumber \\
&\times \exp \left( -i\int d^{2}wd^{2}v\bar{\eta}_{B}\left( w\right)
\mathfrak{S}_{BF}\left( w,v\right) \eta _{F}\left( v\right) \right) .
\end{align}
As it was mentioned earlier, we are interested only in those fundamental Green's functions which have
the general dependence on the fermionic sources such as \eqref{eq a.11}. Therefore, bearing this
thought, the non-quadratic terms in the fermionic sources are dismissed along the calculation. Nevertheless,
in more general correlation functions, further terms, in different order in the fermionic sources,
would be present in the above expressions. Now, by following the same steps and identities presented
previously, one can calculate the remaining nonvanishing terms of the expansion
$ \Phi \left( \eta ,\bar{\eta};m\right) $ without any further difficulty. Therefore, it is not difficult
to one to get the following expression for the equation \eqref{eq a.6.1} until its \textit{n-th} term
\begin{align}
 \Phi \left( \eta ,\bar{\eta};m\right)  =&\bigg\{1+\left( -im\right) \int d^{2}x_{1}d^{2}x_{2}d^{2}
 x_{3}\left[ \bar{\eta}\left( x_{2}\right)  \mathfrak{S}\left( x_{2},x_{1}\right) \mathfrak{S}
 \left( x_{1},x_{3}\right) \eta\left( x_{3}\right) \right]   \notag \\
&+\left( -im\right) ^{2}\int d^{2}x_{1}\ldots d^{2}x_{4}\left[ \bar{\eta}\left( x_{3}\right)\mathfrak{S}\left(
x_{3},x_{1}\right) \left( i\right) \mathfrak{S}\left(x_{1},x_{2}\right)  \mathfrak{S}\left(
x_{2},x_{4}\right) \eta \left( x_{4}\right) \right] \notag \\
&+\ldots+\left( -im\right) ^{n}\int d^{2}x_{1}d^{2}x_{2}\ldots d^{2}x_{n}\Big[\bar{\eta}\left( x_{3}\right)
\mathfrak{S}\left(x_{3},x_{1}\right) \left( i\right) \mathfrak{S}\left(x_{1},x_{2}\right) \notag \\
&\times \ldots\left( i\right) \mathfrak{S}\left( x_{n-2},x_{n-1}\right)  \mathfrak{S}
\left( x_{n-1},x_{n}\right) \eta \left( x_{n}\right) \Big] +\ldots\bigg\}  \notag \\
&\times \exp \left( -i\int d^{2}wd^{2}v\bar{\eta}_{B}\left( w\right)
\mathfrak{S}_{BF}\left( w,v\right) \eta _{F}\left( v\right) \right) . \label{eq 1.11a}
\end{align}%
Since we have already calculated an explicit expression for the quantity $ \Phi \left( \eta ,\bar{\eta};m\right)$,
Eq.\eqref{eq 1.11a}, we are now able to compute the remaining fermionic derivatives present in the expression
\eqref{eq a.11}. Hence, by evaluating the first terms explicitly here, it follows that
\begin{align}
\Phi _{IK}^{\left( 0\right) }\left( x,y\right) &=\bigg.\frac{\delta ^{2}}{\delta \eta _{K}\left( y\right)
\delta \bar{\eta}_{I}\left( x\right) }\exp\left( -i\int d^{2}wd^{2}v\bar{\eta}_{B}\left( w\right) \mathfrak{S}%
_{BD}\left( w,v\right) \eta _{D}\left( v\right) \right) \bigg\vert_{\eta=\bar{\eta}=0}  \notag \\
&=i\mathfrak{S}_{IK}\left( x,y\right) =\exp \left[ -i\int d^{2}uB_{\mu
}\left( u\right) s^{\mu }\left( u,x,y\right) \right] S_{IK}\left( x,y\right),
\end{align}%
where the result \eqref{eq a.7} has been used. We have for the next nonvanishing term in the expansion
\begin{align}
\Phi _{IK}^{\left( 1\right) }\left( x,y\right)  =&\left( -im\right) \bigg.\frac{\delta ^{2}}{\delta
\eta _{K}\left( y\right) \delta \bar{\eta}_{I}\left( x\right) }\left[ \bar{\eta}_{B}\left( x_{2}\right)
\mathfrak{S}_{BD_{1}}\left( x_{2},x_{1}\right)\mathfrak{S}_{D_{1}F}\left( x_{1},x_{3}\right) \eta _{F}
\left( x_{3}\right) \right] \notag \\
&\times \exp \left( -i\int d^{2}wd^{2}v\bar{\eta}_{B}\left( w\right)\mathfrak{S}_{BD}\left( w,v\right)
\eta _{D}\left( v\right) \right) \bigg\vert_{\eta =\bar{\eta}=0}  \notag \\
=&\left( im\right) \int d^{2}x_{1}\left(\mathfrak{S}\left( x,x_{1}\right)
\mathfrak{S}\left( x_{1},y\right) \right)_{IK},
\end{align}%
also, using the result \eqref{eq a.7} combined with some identities and further manipulation, one can
write the last expression as
\begin{equation}
\Phi _{IK}^{\left( 1\right) }\left( x,y\right) =\left( im\right) \exp \left[-i\int d^{2}uB_{\mu }
\left( u\right) s^{\mu }\left( u,x,y\right) \right]\int d^{2}x_{1}\left( S\left( x,x_{1}\right)
S\left( x_{1},y\right) \right) _{IK}.
\end{equation}%
The following terms in the expansion \eqref{eq 1.11a} are evaluated in the same way; moreover, it can be
shown that they have exactly the same structure as the first two ones. Therefore, the complete contribution of them can be casted as
\begin{align}
\Phi _{IK}\left( x,y\right) =&i\exp \left[ -i\int d^{2}uB_{\mu }\left(u\right) s^{\mu }\left( u,x,y\right)
\right] \bigg\{ S_{IK}\left( x,y\right)+m \int d^{2}x_{1}\left( S\left( x,x_{1}\right) S\left(
x_{1},y\right) \right) _{IK}\notag \\
&+m ^{2}\int d^{2}x_{1}d^{2}x_{2}\left( S\left( x,x_{1}\right) S\left( x_{1},x_{2}\right)
S\left( x_{2},y\right) \right) _{IK} +... \notag  \\
&+m^{n}\int d^{2}x_{1}...d^{2}x_{n}\left(S\left( x,x_{1}\right) S\left( x_{1},x_{2}\right) ...S\left(
x_{n-1},x_{n}\right) S\left( x_{n},y\right) \right) _{IK}+...\bigg\}.  \label{eq a.12a}
\end{align}%
It is worth to emphasize that the mass effects surprisingly do not involve the gauge field, as
it can be seen explicitly in the Eq.\eqref{eq a.12a}, its contribution are, in fact, given only in terms of
power of product of the free massless Dirac propagator. Therefore, this feature allows that the gauge
field can be integrated exactly as it is usually performed in the massless case. At last, looking to
the expression of the sum of the product of $m$-dependent terms in \eqref{eq a.12a}, we also can write
the free Dirac propagator at Fourier space, which allows us to evaluate the remaining integrals and
have a better visualization of the final result. For instance, the first couple of terms are easily
written in its momentum representation as
\begin{align*}
\int d^{2}x_{1}\left( S\left( x,x_{1}\right) S\left( x_{1},y\right) \right)_{IK} &=\int \frac{d^{2}p}
{\left( 2\pi \right) ^{2}}\left( \frac{1}{\widehat{p}}\frac{1}{\widehat{p}}\right) _{IK}e^{-ip\left( x-y\right) }, \\
\int d^{2}x_{1}d^{2}x_{2}\left( S\left( x,x_{1}\right) S\left(x_{1},x_{2}\right) S\left( x_{2},y\right)
\right) _{IK} &=\int \frac{d^{2}p}{\left( 2\pi \right) ^{2}}\left( \frac{1}{\widehat{p}}\frac{1}{\widehat{p}}%
\frac{1}{\widehat{p}}\right) _{IK}e^{-ip\left( x-y\right) }.
\end{align*}
The next terms have exactly the same form as these ones, thus it shows that we can sum the fermionic mass contribution
terms in a surprisingly simple way, as a geometric series. Therefore, it follows that the resulting expression,
by taking into account all the massive contribution, it is finally written as
\begin{equation}
\Phi _{IK}\left( x,y\right) =i\exp \left[ -i\int d^{2}uB_{\mu }\left(
u\right) s^{\mu }\left( u,x,y\right) \right] \int \frac{d^{2}p%
}{\left( 2\pi \right) ^{2}}\left[ \frac{1}{%
\widehat{p}-m}\right] _{IK}e^{-ip\left( x-y\right) }.  \label{eq a.13}
\end{equation}%
where we do have now the presence of the free massive Dirac propagator: $S\left( x;m\right) =\int \frac{d^{2}p%
}{\left( 2\pi \right) ^{2}}\left[\frac{1}{\widehat{p}-m}\right] e^{-ipx}$. We have obtained a closed expression
for the fermionic sector of theory, Eq.\eqref{eq a.13}, a result that can easily be expanded to different
Green's functions which may dependent on higher power of the derivatives of the sources $\eta $ and $\bar{\eta}$.
Hence, all the previous results obtained in \cite{14} can be generalized and written in terms of the fermionic
mass, since the only difference is the replacement of a free massless Dirac propagator by a massive one in the quantity
$\Phi _{IK}\left( x,y\right) $, Eq.\eqref{eq a.13}.

Therefore, we can conclude that, by using our proposal, Eq.\eqref{eq a.5c}, instead of the usual approach,
Eqs.\eqref{eq 3.2a} and \eqref{eq 3.2}, which corresponds to a mass insertion scheme into a nonperturbative
massless theory, in particular, we presented the analysis on the Gauged Thirring model, we can still compute
exactly the fundamental Green's functions of a fermionic field theory with a massive fermionic field
interacting with a gauge field in $\left( 1+1\right) $-dimensions, revealing thus the presence of a non-trivial
and interesting structure, and also that exact nonperturbative results in the massive case may be obtained.
We also believe that these results and conclusions can also be applied (or generalized)
to more complicated theories and/or correlation functions.

We shall conclude this section with a brief discussion on the way that the $\theta$-vacuum can be
incorporated in our approach, and its effects on the theory's solution, Eq.\eqref{eq a.13}. As it
is well-known, the $\theta$-vacuum can be studied by adding a term into the generating functional
\eqref{eq a.4} in such a way \cite{20}:
\begin{equation}
\mathscr{Z}\left( J\right) =\int DA_{\mu }D%
\bar{\psi}D\psi \exp \left( i\int d^{2}x\left[ \tilde{\mathcal{L}}+J.\Phi\right] \right) .
\end{equation}
now with a modified Lagrangian density
\begin{align}
\tilde{\mathcal{L}}=\bar{\psi}\left( i\hat{\partial}+\hat{A} -m\right) \psi
-\frac{1}{4e^{2}}F_{\mu \nu }F^{\mu \nu }  +\frac{1}{2g}A^{\mu }A_{\mu }
-\frac{1}{2\xi }\left( \partial _{\mu }A^{\mu }\right) ^{2}-\frac{e\theta}{4\pi}\epsilon_{\mu\nu}F^{\mu\nu}. \label{eq 1.3}
\end{align}%
For simplicity we shall take the instanton number equal to zero in order to preserve the expression
of the solution of the external field fermionic propagator \cite{22}. In principle, the
$\theta$-term on \eqref{eq 1.3} can be eliminated by means of a finite chiral rotation,
\begin{equation}
\psi \left( x\right) =e^{ i\gamma _{5}\alpha  } \psi \left( x\right) , \quad \bar{\psi} \left( x\right)
=\bar{\psi}\left( x\right) e ^{ i\gamma _{5}\alpha } , \label{eq c}
\end{equation}%
since it originates a Jacobian in the fermionic measure
\begin{equation}
J_{C}=\exp\left[\frac{ie\alpha}{2\pi}\int d^{2}z \epsilon_{\mu\nu}F^{\mu\nu}\right],
\end{equation}
and by choosing $\alpha=\theta/2$ it cancels the $\theta$-term in the equation \eqref{eq 1.3}. However,
the fermionic sources and mass terms are also changed by the chiral rotation \eqref{eq c} as,
\begin{equation}
m\bar{\psi}\psi  +\bar{\eta}\psi+\bar{\psi}\eta \rightarrow m\bar{\psi}e^{i\theta \gamma_{5}}\psi +\bar{\eta}e^{\frac{i\theta \gamma_{5}}{2}}\psi+\bar{\psi}e^{\frac{i\theta \gamma_{5}}{2}}\eta .
\end{equation}
However, one can easily show that the above mass term can be written exactly as it was casted in our
previous discussion in the generating functional Eq.\eqref{eq a.5c}, i.e., in terms of functional derivatives.
Therefore, in our discussion, the only $\theta$-vacuum contribution is in the fermionic source terms.
Finally, we can obtain that the remaining $\theta$-vacuum contribution in the expression of the
generating functional \eqref{eq a.5c} is
\begin{equation}
\exp \left( -i\int d^{2}xd^{2}y\bar{\eta}_{B}\left( x\right)e^{\frac{i\theta \gamma_{5}}{2}}
\mathfrak{S}_{BD}\left( x,y;A+C\right)e^{\frac{i\theta \gamma_{5}}{2}} \eta _{D}\left( y\right) \right) .
\end{equation}
Moreover, we can make use of the explicit solution of the external field fermionic propagator \eqref{eq a.7}
to show that the above contribution of the $\theta$-vacuum is canceled out by evaluating explicitly the product:
$[\bar{\eta}e^{\frac{i\theta \gamma_{5}}{2}} \mathfrak{S}\left( x,y;B\right)e^{\frac{i\theta
\gamma_{5}}{2}}\eta ]=[\bar{\eta}\mathfrak{S}\left( x,y;B\right)\eta ]$. Therefore, we see that any reference
to the $\theta$-vacuum is eliminated from the theory's solution by making a finite chiral rotation, at least
at instanton number equals to zero.

\section{Path-Integral Bosonization}

\label{sec:1}

As a matter of complementarity on our analysis of mass effects in a two-dimensional field theory, we now discuss
the MGTM by a different perspective, the bosonization framework. This framework is rather suitable and interesting
for the analysis of massive two-dimensional theories. Our interest in developing the bosonized description of the model's
consists mainly as a part of a subsequent investigation of two-dimensional theories in curved
spacetime. Nevertheless, we start the analysis heading back to the Lagrangian density Eq.\eqref{eq a.1}:
\begin{equation}
\mathcal{L}=\bar{\psi}\left( i\gamma .\partial +\gamma .A\right) \psi -m\bar{\psi}\psi +\frac{1}{2g}\left( A_{\mu }
-\partial _{\mu }\theta \right) ^{2}-\frac{1}{4e^{2}}F_{\mu \nu }F^{\mu \nu }.  \label{eq 1}
\end{equation}%
Moreover, it is possible to write the gauge field, in two-dimensions, decomposed into its longitudinal
and transverse components as:
\begin{equation}
A_{\mu }\left( x\right) =\partial _{\mu }\rho \left( x\right) -\tilde{\partial}_{\mu }\phi \left( x\right) .  \label{eq 1.2}
\end{equation}%
As we have discussed previously, a suitable gauge condition for the MGTM is the $R_{\xi }-$%
gauge \eqref{eq a.3.1}. With this gauge condition, it is clear that the gauge field takes the form:
\begin{equation}
\square \rho \left( x\right) +\frac{\xi }{g}\theta \left( x\right) =0. \label{eq 1.4}
\end{equation}
On the other hand, we have that from the above decomposition \eqref{eq 1.2} an interaction term appeared
between the fermionic fields with the scalar ones ($\rho$ and $\phi $). Thus, in order to canceling out
this coupling between the fermionic with the scalar fields $\rho$, we may perform a dressing into the
fermionic variables \cite{23}, namely,
\begin{align}
\psi \left( x\right) =e^{ i \rho \left( x\right) -i\gamma _{5}\phi \left( x\right) } \chi \left( x\right) ,
\quad \bar{\psi} \left( x\right) =\bar{\chi}\left( x\right) e ^{ -i \rho \left(x\right)
-i\gamma _{5}\phi \left( x\right) } ,  \label{eq 1.5}
\end{align}%
which correspond (in the path-integral framework) to the bosonization realization in the operator
approach. Hence, it follows that, under the change of variables: \eqref{eq 1.2} and \eqref{eq 1.5},
and also by eliminating the $\theta$-field through the identity \eqref{eq 1.4}, one can write the Lagrangian \eqref{eq 1} as
\begin{equation}
\mathcal{L}_{eff}=\bar{\chi}i\widehat{\partial }\chi -m\bar{\chi}e^{-2i\gamma _{5}\phi }\chi +\frac{1}{2g}
\left( 1+\frac{g}{\xi }\square ^{2}\right) ^{2}\left(\partial _{\mu }\rho \right) ^{2}
+\frac{1}{2g}\left( \tilde{\partial}_{\mu }\phi \right)^{2} +\frac{1}{2e^{2}}\phi \square ^{2}\phi.
\end{equation}
However, we still have a remaining interaction between the fermionic and $\phi $ fields in the above Lagrangian.
Such interaction can be solved by an appropriated perturbative mass expansion. But, before performing this mass
expansion, we need to pay some attention to an important consideration regarding the bosonization in the
path-integral framework. In such framework, the fields measure of the original transition-amplitude
changes its form under the transformations \eqref{eq 1.2} and \eqref{eq 1.5}:
\begin{equation}
DA_{\mu }=J_{A}D\rho D\phi ,\quad D\bar{\psi}D\psi =J_{F}D\bar{\chi}D\chi , 
\end{equation}%
with $J_{A}$ equals to a constant, that can be absorbed into the normalization constant $N$ \cite{20}.
Moreover, the fermionic Jacobian $J_{F}$ has a nontrivial expression due to the axial anomaly \cite{10},
and because of the Abelian character of the MGTM its evaluation is direct, resulting into:%
\begin{equation}
J_{F}=\exp \left[ -i\frac{1}{2\pi }\int d^{2}x\left( \partial _{\mu }\phi
\right) \left( \partial ^{\mu }\phi \right) \right] . \label{eq 1.7a}
\end{equation}
Therefore, combining the facts: all the variable changing, subsequent manipulation, one can obtain the
following transition-amplitude expression:%
\begin{align}
Z=&N^{\prime }\int D\bar{\chi}D\chi D\phi D\rho \exp \Big[i\int d^{2}x \Big(\bar{\chi}
\left(i\widehat{\partial }-m e^{-2i\gamma _{5}\phi}\right)\chi \notag\\
&+\frac{1}{2e^{2}}\phi \square \left[ \square +M^{2}\right] \phi -\frac{1}{2g}%
\left( 1+\frac{g}{\xi }\square \right) ^{2}\rho \square \rho \Big)\Big].   \label{eq 1.8}
\end{align}%
Here it was defined: $M^{2}=e^{2}\left( \frac{1}{\pi }+\frac{1}{g}\right) $. As we can see in the
expression \eqref{eq 1.8}, the scalar field $\rho $ is completely decoupled from the other fields.
Hence, its contribution can be absorbed into the normalization constant. Now, we can proceed into
the perturbative expansion of the fermion mass \cite{20}, to then evaluate the fermionic and bosonic
Wightman's functions. To accomplish that, we can write the transition-amplitude \eqref{eq 1.8} in the form:
\begin{align}
Z =& N^{\prime \prime }\int D\bar{\chi}D\chi D\phi   \exp \left[ i\int d^{2}x\left( \bar{\chi}
i\widehat{\partial }\chi +\frac{1}{2e^{2}}\phi \square \left[ \square +M^{2}\right] \phi \right) \right]  \notag \\
& \times \underset{k=0}{\overset{\infty }{\sum }}\frac{\left( -im\right) ^{k}}{k!}\underset{i=1}
{\overset{k}{\prod }}\int d^{2}x_{i}\bar{\chi}\left( x_{i}\right) e^{-2i\gamma _{5}\phi \left( x_{i}\right) }
\chi \left( x_{i}\right) , \label{eq 1.9}
\end{align}%
whose expression immediately reads:%
\begin{equation}
Z=\underset{k=0}{\overset{\infty }{\sum }}\frac{\left( -im\right) ^{k}%
}{k!}\left\langle \underset{i=1}{\overset{k}{\prod }}\int d^{2}x_{i}\bar{\chi}%
\left( x_{i}\right) e^{-2i\gamma _{5}\phi \left( x_{i}\right) }\chi \left(
x_{i}\right) \right\rangle _{0}.  \label{eq 1.11}
\end{equation}%
Here $\left\langle {\quad }\right\rangle _{0}$ stands for the vacuum expectation value (vev) of
a set of operators in a system of massless free fermion and massive free scalar fields. In order
to evaluate the expression \eqref{eq 1.11} we need to separate the bosonic and fermionic fields
on the argument of the vev. For this purpose, we can rewrite the $\gamma_{5}$ factor from the
exponential of \eqref{eq 1.11} in such a way that permits one to get
\begin{align}
Z =&\underset{n=0}{\overset{\infty }{\sum }}\frac{\left( -im\right) ^{2n}}{\left( n!\right) ^{2}}
\int \left( \underset{k=1}{\overset{n}{\prod }}d^{2}x_{k}d^{2}y_{k}\right)  \left\langle \exp
\left( -2i\underset{j}{\sum }\left( \phi \left( x_{j}\right) -\phi \left( y_{j}\right) \right) \right)
\right\rangle _{0}^{bos} \notag \\
&\times \left\langle \underset{i=1}{\overset{n}{\prod }}\bar{\chi}\left(x_{i}\right) \frac{\left(
1+\gamma _{5}\right) }{2}\chi \left( x_{i}\right)\bar{\chi}\left( y_{i}\right)
\frac{\left( 1-\gamma _{5}\right) }{2}\chi\left( y_{i}\right) \right\rangle _{0}^{f}.  \label{eq 1.12}
\end{align}%
The simplest term to compute in \eqref{eq 1.12} is the fermionic contribution; where the fermionic
Wightman function is just the massless free fermion propagator:
\begin{equation}
S_{F}\left( x\right) =-\frac{1}{2\pi }\frac{\gamma ^{\mu }x_{\mu }}{x^{2}}. \label{eq 1.20}
\end{equation}%
To evaluate the fermionic part, we shall decompose the spinors in their components, allowing us
thus obtain the result \cite{1}:
\begin{equation}
\left\langle \underset{i=1}{\overset{n}{\prod }}\bar{\chi}_{1}\left(x_{i}\right) \chi _{1}\left(
x_{i}\right) \bar{\chi}_{2}\left( y_{i}\right) \chi _{2}\left( y_{i}\right) \right\rangle _{0}^{f}
=\frac{1}{\left( 2\pi i\right) ^{2n}}\frac{\overset{n}{\underset{i>j}{\prod }}\left(c^{2}\left\vert
x_{i}-x_{j}\right\vert ^{2}\left\vert y_{i}-y_{j}\right\vert^{2}\right) }{\underset{i,j}{\overset{n}
{\prod }}\left( c\left\vert x_{i}-y_{j}\right\vert ^{2}\right) }, \label{eq 1.15}
\end{equation}%
where $c=e^{-\gamma }$, with $\gamma $ the Euler-Mascheroni constant. Furthermore, to calculate the
bosonic contribution, we must first evaluate the scalar Wightman function. We can define it as it follows:
\begin{equation}
\frac{1}{e^{2}}\square \left[ \square +M^{2}\right] \Delta \left( x\right)
=-\delta ^{\left( 2\right) }\left( x\right) , \notag
\end{equation}%
its solution is readily obtained:
\begin{equation}
\Delta \left( x\right) =\lambda ^{2}\left( -\frac{1}{4\pi }\ln \left(
M^{2}c^{2}x^{2}\right) -\frac{1}{2\pi }K_{0}\left( \sqrt{M^{2}x^{2}}\right) \right) ;  \label{eq 1.7}
\end{equation}%
with: $\lambda ^{2}=\frac{\pi }{1+\frac{\pi }{g}}$, and $K_{0}\left(z\right) $ the second-class modified
Bessel's function. In possess of these elements, one is able to obtain the well-known result for the scalar vev \cite{1}:
\begin{align}
\left\langle \exp \left( -2i\underset{j}{\sum }\left( \phi \left(x_{j}\right) -\phi \left( y_{j}\right)
\right) \right) \right\rangle _{0}^{bos}  =\exp \Bigg[4\underset{i>j}{\sum }\Big(\Delta \left(x_{i}-x_{j}\right) +\Delta \left( y_{i}-y_{j}\right)
-\Delta \left(x_{i}-y_{j}\right) \Big)\Bigg],  \label{eq 1.13}
\end{align}%
also, substituting $\Delta \left( x\right) $ by its explicit expression \eqref{eq 1.7} in the last equation,
it yields to the expression:
\begin{align}
&\left\langle \exp \left( -2i\underset{j}{\sum }\left( \phi \left( x_{j}\right) -\phi \left( y_{j}\right)
\right) \right) \right\rangle _{0}^{bos}  =\left[ Mc\right] ^{\frac{2n\lambda ^{2}}{\pi }}\frac{\overset{n}
{\underset{i>j}{\prod }}\left(\left\vert x_{i}-x_{j}\right\vert^{2} \left\vert y_{i}-y_{j}\right\vert^{2}
\right)^{-\frac{\lambda ^{2}}{\pi }}}{\underset{i,j}{\overset{n}{\prod }}\left\vert x_{i}-y_{j}\right\vert ^{-
\frac{2\lambda ^{2}}{\pi }}} \notag \\
& \times \exp \bigg[ \frac{2\lambda ^{2}}{\pi }\underset{i>j}{\sum }  \bigg(K_{0}\left( M;\left\vert
x_{i}-y_{j}\right\vert \right) -K_{0}\left( M;\left\vert x_{i}-x_{j}\right\vert \right)
-K_{0}\left(M;\left\vert y_{i}-y_{j}\right\vert \right)\bigg) \bigg] .\label{eq 1.14a}
\end{align}%
Therefore, with the previous results: \eqref{eq 1.15} and \eqref{eq 1.14a}, we can finally get the expression for the
transition-amplitude \eqref{eq 1.12}, which is then written as:
\begin{align}
 Z=&\underset{n=0}{\overset{\infty }{\sum }}\frac{1}{\left( n!\right) ^{2}}\left( \frac{m}{2\pi c}\left[ Mc\right]
 ^{\frac{\lambda ^{2}}{\pi }}\right) ^{2n}  \int \left( \underset{k=1}{\overset{n}{\prod }}d^{2}x_{k}d^{2}y_{k}\right) \frac{\overset{n}{\underset{i>j}{\prod }}\left(\left\vert x_{i}-x_{j}\right\vert ^{2}\left\vert y_{i}-y_{j}\right
 \vert ^{2}\right)^{\left( 1-\frac{\lambda ^{2}}{\pi }\right) }}{\underset{i,j}{\overset{n}{\prod }}\left\vert
x_{i}-y_{j}\right\vert ^{2\left( 1-\frac{\lambda ^{2}}{\pi }\right) }} \notag \\
&\times\exp \bigg[ \frac{2\lambda ^{2}}{\pi }\underset{i>j}{\sum }  \bigg(K_{0}\left( M;\left\vert x_{i}-y_{j}
\right\vert \right)-K_{0}\left( M;\left\vert x_{i}-x_{j}\right\vert \right) -K_{0}\left(M;\left\vert y_{i}-y_{j}
\right\vert \right)  \bigg) \bigg] .\label{eq 1.16}
\end{align}%
Although the equivalence between the MGTM with the MSM and MTM in the bosonized transition-amplitude
seems almost trivial, it is worth to mention some features of the transition-amplitude \eqref{eq 1.16}.
Differently from the equivalence of the MGTM in the quantum level \cite{14}, where the Green's functions were
analyzed in face of the proper limits: $g\rightarrow\infty$ and $e^{2}\rightarrow \infty$, respectively, and
also by a careful discussion about the gauge symmetry, one can show here the convergence of the expression
\eqref{eq 1.16} to the respective models \cite{2,20} through the same limits.

As it was mentioned earlier, we intend to extend the results of the Eqs.\eqref{eq 1.8} and \eqref{eq 1.16}
to the case of curved spacetime, with a particular interest in investigating how its solutions,
thermodynamical and other important quantities, are affected by the presence of a gravitational background.

\section{Concluding Remarks}

\label{sec:3}

We have investigated the behavior of the solutions of the Gauged Thirring model when the fermionic field is given a mass.
The study consisted in analyzing which properties of the massless theory are preserved, or broken, in this generalization.
To implement this idea, we have used standard functional techniques in the path-integral framework in order to apply
the mass insertions into the generating functional of the massless theory. Our major interest in this analysis relied
specifically in determining whether the nonperturbative nature of the massless solutions (fermionic determinant)
was preserved in this generalization. Nevertheless, for our great surprise, the fermionic mass contributes to the Green's functions
in a highly surprisingly simple way, leaving therefore the original nonperturbative nature of the massless results
still intact in the massive theory. As a first analysis, we believe that we have supplied sufficient details about our
calculation and results on the mass contribution to the theory's fundamental quantities. We were mainly motivated in
choosing the Gauged Thirring model as a laboratory to investigate this idea, once it possesses a non-trivial structure
and also by containing the Schwinger and Thirring models as particular cases when the proper limits are taken
at both classical and quantum level.

Although several papers were devoted in studying the subject of massive two-dimensional field theories, we
believe that there are still many interesting issues and phenomena untreated so far. We are conscious that the
claims made here deserve more attention, and that they should be analyzed in further investigations. One can even
ask whether these claims hold when it is considered thermodynamical quantities of the theory, or even whether
a non-trivial structure of spacetime affects the nature of the solution of these exactly soluble models.
Hence, strongly motivated by the study on the theory's behavior in face of a classical curved background, we have also
derived here the theory's bosonized effective action and its generating functional as well. This approach provides an
interesting framework in studying the nonperturbative solution of a massive fermionic theory. Furthermore,
it also shows to be suitable, a priori, to obtain information about the effects of geometry on the theory's
phenomena \cite{32}. Nevertheless, work on these aspects will be presented elsewhere.

\section*{Acknowledgment}

RB thanks CNPq and FAPESP for full support and BMP thanks CNPq and CAPES for
partial support.

\bibliographystyle{elsarticle-num}

\end{document}